# Sufficient condition for validity of quantum adiabatic theorem

Yong Tao[†]

School of Economics and Business Administration, Chongqing University, Chongqing 400044, China

**Abstract:** In this paper, we attempt to give a sufficient condition of guaranteeing the validity of the proof of the quantum adiabatic theorem. The new sufficient condition can clearly remove the inconsistency and the counterexample of the quantum adiabatic theorem pointed out by Marzlin and Sanders [Phys. Rev. Lett. 93, 160408, (2004)].

## 1. Introduction

The quantum adiabatic theorem (QAT) [1-4] is one of the basic results in quantum physics. The role of the QAT in the study of slowly varying quantum mechanical system spans a vast array of fields and applications, such as quantum field [5], Berry phase [6], and adiabatic quantum computation [7]. However, recently, the validity of the application of the QAT had been doubted in reference [8], where Marzlin and Sanders (MS) pointed out an inconsistency. This inconsistency led to extensive discussions from physical circles [9-18]. Although there are many different viewpoints as for the inconsistency, there seems to be a general agreement that the origin is due to the insufficient conditions for the QAT.

More recently, we have summarized these different viewpoints of studying the inconsistency of the QAT through the reference [19] where we have noticed that there are essentially two different types of inconsistencies of the QAT in reference [8], that is, MS inconsistency and MS counterexample. Most importantly, these two types are often confused as one [19]. That is just the reason why there are many different viewpoints as for the inconsistency of the QAT. Our previous study [19] shows that removing the MS counterexample depends on the convergence of the Schrödinger integral equation in the adiabatic limit (References [12,14,15,16,17,18] refer to this point), and that resolving the MS inconsistency depends on the convergence of the Schrödinger differential equation in the adiabatic limit (References [10,13] refer to this point). Nevertheless, a lot of references always pay attention to the MS counterexample rather than the MS inconsistency. In reference [19] we point out that the MS inconsistency is very important: Resolving the MS inconsistency would give rise to the necessity of integral formalism. For example, we can only reach a complete QAT through Schrödinger integral equation rather than Schrödinger differential equation.

Our main purpose of this paper is to provide a new sufficient condition to rigorously prove the complete QAT through the Schrödinger integral equation. Later, we shall find that the new sufficient condition can clearly remove the MS inconsistency and the MS counterexample.

---

[†] Corresponding author.
E-mail address: taoyingyong2007@yahoo.com.cn

## 2. Two types of the proof of the quantum adiabatic theorem

Firstly, we need to emphasize that our derivation to the QAT is a more general proof in comparison with the standard proof (e.g., Messiah's proof [4]). That is because the starting point for our discussion is the Schrödinger equation[1]

$$i\frac{1}{T}\frac{\partial}{\partial s}|\psi(sT)\rangle = H(sT)|\psi(sT)\rangle, \quad s \in [0,1] \quad (1)$$

where the Hamiltonian $H(sT)$ not only depends on the scaled dimensionless time $s$ but also on the total evolution time $T$.

As a result, the instantaneous eigenstates and energies are the solutions of a family of Hamiltonian equations parameterized by $s$ and $T$,

$$H(sT)|n(sT)\rangle = E_n(sT)|n(sT)\rangle.$$

The QAT states that as $T \to \infty$ [2], the solution to the Schrödinger equation (1) approaches

$$|\psi(sT)\rangle_A = \exp\left[-iT\int_0^s E_n(s'T)ds'\right]\exp\left[-\int_0^s \langle n(s'T)|\frac{\partial}{\partial s'}|n(s'T)\rangle ds'\right]|n(sT)\rangle. \quad (2)$$

In general, to arrive at the QAT, there are two types of the proof of the QAT as follows:

(i) Proof of differential formalism: $\lim_{T \to \infty}\frac{\partial}{\partial s}|\psi(sT)\rangle = \frac{\partial}{\partial s}|\psi(sT)\rangle_A$.

(ii) Proof of integral formalism: $\lim_{T \to \infty}|\psi(sT)\rangle = |\psi(sT)\rangle_A$.

Recently, nevertheless, Wu and Yang [10] have pointed out that the proof of differential formalism (i) is invalid and meanwhile may give rise to the MS inconsistency. More importantly, up to now, those rigorous proofs of the QAT are, essentially, all based on the proof of integral formalism (ii) [3,4,22,23,24]; but this key fact is always neglected by physical circles. Unfortunately, if we neglect this fact and hence use the proof of differential formalism (i) to derive the QAT, we would reach the MS inconsistency [10,19].

Clearly, to guarantee the validity of the proof of integral formalism (ii), there needs a sufficient condition. Unfortunately, the traditional adiabatic condition [25], denoted by

$$\langle n(t)|\frac{\partial}{\partial t}|m(t)\rangle = 0 \quad (n \neq m),$$

is insufficient and hence can not guarantee the validity of the proof of integral formalism (ii)[3]. Later, we shall notice that neglecting this fact would lead to the MS counterexample.

In next section, we shall give a new sufficient condition of guaranteeing the validity of the proof of integral formalism (ii).

---

[1] We have used the scaled dimensionless time variable $s = \dfrac{t}{T}$ in the Schrödinger equation.

[2] $T \to \infty$ denotes the adiabatic limit. For the case of non-adiabatic evolution, i.e., $T < \infty$, sees reference [20]. Interestingly, the non-adiabatic effect has been noticed by a recent experiment [21].

[3] In fact, the traditional adiabatic condition, denoted by $\langle n(t)|\dfrac{\partial}{\partial t}|m(t)\rangle = 0 \; (n \neq m)$, is related to the proof of differential formalism (i) [25] and hence is invalid [10].

## 3. New sufficient condition of the quantum adiabatic theorem

Before proceeding to present the new sufficient condition of the QAT, we need to give an exact expression of Berry phase to guarantee the mathematical stringency of the following discussion.

Strictly speaking, mathematically, there may be, in the adiabatic limit $T \to \infty$, that $\lim_{T \to \infty} |n(sT)\rangle$ does not exist, and that $\lim_{T \to \infty} \frac{\partial}{\partial s} |n(sT)\rangle \neq \frac{\partial}{\partial s} \lim_{T \to \infty} |n(sT)\rangle$. This case implies that $|n(sT)\rangle$ is singular for $T \to \infty$. In such singular situation, the Berry phase, which is denoted by $i \oint \lim_{T \to \infty} \langle n(sT) | d | n(sT) \rangle$ [26], is not meaningful. Therefore, if we want to require that the Berry phase makes good sense, then the theorem 3.1 given as follow must hold.

**Theorem 3.1**[4]. If $\lim_{T \to \infty} |n(sT)\rangle = |n(s)\rangle$ and if $\lim_{T \to \infty} \frac{\partial}{\partial s} |n(sT)\rangle = \frac{\partial}{\partial s} \lim_{T \to \infty} |n(sT)\rangle$ [27], then the Berry phase is uniquely determined by

$$\gamma(C) = \lim_{T \to \infty} i \oint \langle n(sT) | d | n(sT) \rangle = i \oint \lim_{T \to \infty} \langle n(sT) | d | n(sT) \rangle = i \oint \langle n(s) | d | n(s) \rangle.$$

*Proof.* Using the bounded convergence theorem of Lebesgue [28], the proof is complete. □

In addition to an exact expression of the Berry phase, the singular limit in the Schrödinger equation (1) may also affect the validity of the proof of the QAT, for example, the singular limit[5] $T \to \infty$ would eliminate the derivative of the wave function $\frac{\partial}{\partial s} |\psi(sT)\rangle$. To remove the singular limit, an adiabatic transformation denoted by

$$|\psi(sT)\rangle = A(sT) |\phi(sT)\rangle \qquad (3)$$

need to be constructed,

where, $A(sT) = \sum_n \exp\left[-iT \int_0^s E_n(s'T) ds'\right] |n(sT)\rangle\langle n(0)|$.

Substitution of the equation (3) into the Schrödinger equation (1) yields

$$\frac{\partial}{\partial s} |\phi(sT)\rangle = -K_T(s) |\phi(sT)\rangle, \qquad (4)$$

where,

$$K_T(s) = \sum_j \sum_k \exp\left[iT \int_0^s E_j(s'T) - E_k(s'T) ds'\right] \langle j(sT) | \frac{\partial}{\partial s} | k(sT)\rangle | j(0)\rangle\langle k(0)|. \qquad (5)$$

It is carefully noticed that $K_T(s)$ is an ordinary operator; this is important, because it implies that the solution of Schrödinger differential equation (4) is an ordinary exponential rather than a time-ordered exponential.

---

[4] $|n(s)\rangle$ is independent of the total evolution time $T$.

[5] The importance of the singular limit has been emphasized by Michael Berry in his paper: Singular Limits. Physics Today 55, 10 (2002)

Clearly, Schrödinger differential equation (4) has got ride of the dilemma of singular limit $T \to \infty$. However, we can not reach the complete QAT through Schrödinger differential equation (4) since Wu and Yang have pointed out that the proof of differential formalism (i) based on the equation (4) would lead to the MS inconsistency [10].

Therefore, our starting point for the proof of the QAT is the integral formulation of the equation (4), that is,

$$|\phi(sT)\rangle = |\phi(0)\rangle - \int_0^s K_T(s')|\phi(s'T)\rangle ds'. \qquad (6)$$

In spirit of the reference [19], if we want to prove the QAT, we only need to prove that

$$\lim_{T \to \infty}|\phi(sT)\rangle = \lim_{T \to \infty} \exp\left[-\int_0^s \langle m(s'T)|\frac{\partial}{\partial s'}|m(s'T)\rangle ds'\right]|m(0)\rangle$$
$$= \exp\left[-\int_0^s \langle m(s')|\frac{\partial}{\partial s'}|m(s')\rangle ds'\right]|m(0)\rangle \qquad (7)$$

If we take $|\phi(sT)\rangle = W_T(s)|\phi(0)\rangle$, then our main result of this paper (i.e., the sufficient condition of the QAT) is presented as follow:

***Theorem 3.2.*** If $W_T(s)$ satisfies the following four conditions (a)-(d), then we can prove

$$\lim_{T \to \infty} W_T(s) = \sum_k \exp\left[-\int_0^s \langle k(s')|\frac{\partial}{\partial s'}|k(s')\rangle ds'\right]|k(0)\rangle\langle k(0)|. \qquad (8)$$

(a) The Schrödinger integral equation reads

$$W_T(s) = I - \int_0^s K_T(s')W_T(s')ds', \qquad (9)$$

where $K_T(s)$ is defined by the equation (5).

(b) For an arbitrary eigenstate $|n(sT)\rangle$, there have

$$\lim_{T \to \infty}|n(sT)\rangle = |n(s)\rangle \text{ and } \lim_{T \to \infty}\frac{\partial}{\partial s}|n(sT)\rangle = \frac{\partial}{\partial s}\lim_{T \to \infty}|n(sT)\rangle.$$

(c) $\langle j(sT)|\frac{\partial}{\partial s}|k(sT)\rangle \neq 0$. $(j \neq k)$

(d) $\lim_{T \to \infty} \int_0^s \exp\left[iT\int_0^{s'} E_j(\sigma T) - E_k(\sigma T)d\sigma\right]ds' = 0$ $(j \neq k)$ as for any $s \in [0,1]$.

Clearly, the equation (8) is equivalent to the equation (7). In next section, we shall prove the theorem 3.2.

## 4. Proof of the quantum adiabatic theorem

Before proceeding to prove the theorem 3.2, we need to introduce two lemmas.

**Lemma 4.1.** If $\lim_{T\to\infty} \int_0^s \exp\left[iT\int_0^{s'} E_j(\sigma T) - E_k(\sigma T) d\sigma\right] ds' = 0$ $(j \neq k)$ as for any $s \in [0,1]$, then we can prove

$$\lim_{T\to\infty} \int_0^s \exp\left[iT\int_0^{s'} E_j(\sigma T) - E_k(\sigma T) d\sigma\right] f(s') ds' = 0 \quad (j \neq k),$$

where $f(s)$ is any integrable function on the interval $[0,1]$.

*Proof.* Using the general Riemann-Lebesgue lemma [29], then we note that the lemma 4.1 holds. □

**Lemma 4.2.** If the lemma 4.1 holds, then, for any integrable function $f_T(s)$ which is uniform convergence to $f(s)$, i.e., $\lim_{T\to\infty} \sup_{s\in[0,1]} |f_T(s) - f(s)| = 0$, there would hold

$$\lim_{T\to\infty} \int_0^s \exp\left[iT\int_0^{s'} E_j(\sigma T) - E_k(\sigma T) d\sigma\right] f_T(s') ds'$$
$$= \lim_{T\to\infty} \int_0^s \exp\left[iT\int_0^{s'} E_j(\sigma T) - E_k(\sigma T) d\sigma\right] f(s') ds' = 0 \quad (j \neq k)$$

*Proof.* Let $g_T(s) = \exp\left[iT\int_0^s E_j(\sigma T) - E_k(\sigma T) d\sigma\right]$ $(j \neq k)$, then there holds

$$\left|\int_0^s g_T(s') f_T(s') ds' - \int_0^s g_T(s') f(s') ds'\right| \leq \int_0^s |f_T(s') - f(s')| ds' \leq \sup_{s'\in[0,s]} |f_T(s') - f(s')| s.$$

Using $\lim_{T\to\infty} \sup_{s\in[0,1]} |f_T(s) - f(s)| = 0$, we note that the lemma 4.2 holds. □

Now, we start to prove the theorem 3.2.

*Proof.* Clearly, the equation (9) can be written in the form:

$$W_T(s) = \exp\left[-\int_0^s K_T(s') ds'\right]. \qquad (10)$$

In section 3, we have pointed out that $K_T(s)$ is an ordinary operator. Hence, the equation (10) is an ordinary exponential rather than a time-ordered exponential.

Firstly, we write $K_T(s)$ in the form:

$$K_T(s) = K_T^{(1)}(s) + K_T^{(2)}(s), \qquad (11)$$

where $K_T^{(1)}(s) = \sum_k \langle k(sT)| \frac{\partial}{\partial s} |k(sT)\rangle |k(0)\rangle\langle k(0)|$,

and $K_T^{(2)}(s) = \sum_{j \neq k} \exp\left[ iT \int_0^s E_j(s'T) - E_k(s'T) ds' \right] \langle j(sT)| \frac{\partial}{\partial s} |k(sT)\rangle |j(0)\rangle\langle k(0)|$.

On the one hand, by using the lemma 4.2, the conditions (b) and (d) would guarantee

$$\lim_{T \to \infty} \int_0^s K_T^{(2)}(s') ds' = 0. \qquad (12)$$

On the other hand, by using the bounded convergence theorem of Lebesgue [28], the condition (b) would guarantee

$$\lim_{T \to \infty} \int_0^s K_T^{(1)}(s') ds' = \int_0^s \lim_{T \to \infty} K_T^{(1)}(s') ds' = \sum_k \int_0^s \langle k(s')| \frac{\partial}{\partial s'} |k(s')\rangle |k(0)\rangle\langle k(0)| ds'. \qquad (13)$$

Therefore, substitution of equations (12) and (13) into equation (10), in the adiabatic limit $T \to \infty$, yields

$$\lim_{T \to \infty} W_T(s) = \exp\left[ -\lim_{T \to \infty} \int_0^s K_T(s') ds' \right] = \sum_k \exp\left[ -\int_0^s \langle k(s')| \frac{\partial}{\partial s'} |k(s')\rangle ds' \right] |k(0)\rangle\langle k(0)|.$$

The proof is complete. □

Clearly, the theorem 3.2 shows that $W_T(s)$, in the adiabatic limit $T \to \infty$, satisfies the Schrödinger integral equation (9). Moreover, it is carefully noticed that if we use the equation $\langle j(sT)| \frac{\partial}{\partial s} |k(sT)\rangle = 0 \ (j \neq k)$ rather than the condition (d), then the equation (12) still holds so that the theorem 3.2 holds. Unfortunately, the equation $\langle j(sT)| \frac{\partial}{\partial s} |k(sT)\rangle = 0 \ (j \neq k)$ would give rise to the MS inconsistency [10,19]. However, the condition (c) of theorem 3.2 has removed this difficulty.

Additionally, it is easy to check that the MS counterexample presented by the reference [8] does not fulfill the condition (d) of theorem 3.2. That means that the condition (d) have clearly ruled out the MS counterexample.

## 5. Origin of the inconsistency of the quantum adiabatic theorem

In this section, we attempt to rigorously prove that if $\langle j(sT)| \frac{\partial}{\partial s} |k(sT)\rangle \neq 0 \ (j \neq k)$, then $W_T(s)$ denoted by the equation (8), in the adiabatic limit $T \to \infty$, does not satisfy the Schrödinger differential equation (4), i.e.,

$$\frac{\partial}{\partial s}W_T(s) = -K_T(s)W_T(s). \qquad (14)$$

*Proof.* First, the function structure of $K_T^{(2)}(s)$ shows that $\lim_{T\to\infty} K_T(s)$ does not exist because of the existence of the exponential factor $\exp\left[iT\int_0^s E_j(s'T) - E_k(s'T)ds'\right]$. Thus, by using the equation (14) we easily note that $\lim_{T\to\infty} \frac{\partial}{\partial s}W_T(s)$ also does not exist.

Second, the theorem 3.2 has shown that the limit $\lim_{T\to\infty} W_T(s)$ exists; meanwhile, the condition (b) can guarantee that $\lim_{T\to\infty} W_T(s)$ is differentiable. That means that $\frac{\partial}{\partial s}\lim_{T\to\infty} W_T(s)$ exists.

As a result, we obtain (through comparing above two points)

$$\lim_{T\to\infty} \frac{\partial}{\partial s}W_T(s) \neq \frac{\partial}{\partial s}\lim_{T\to\infty} W_T(s). \qquad (15)$$

Nevertheless, reference [19] has shown that $W_T(s)$, in the adiabatic limit $T\to\infty$, satisfies the Schrödinger differential equation (14) if and only if

$$\lim_{T\to\infty} \frac{\partial}{\partial s}W_T(s) = \frac{\partial}{\partial s}\lim_{T\to\infty} W_T(s). \qquad (16)$$

Clearly, the equation (16) is contrary to the inequality (15).

The proof is complete. □

Conversely, if $W_T(s)$, in the adiabatic limit $T\to\infty$, satisfies the Schrödinger differential equation (14), then there will hold $\langle j(sT)|\frac{\partial}{\partial s}|k(sT)\rangle = 0 \ (j\neq k)$ which would give rise to the MS inconsistency [10,19]. Nevertheless, in reference [8], MS require that $\lim_{T\to\infty} W_T(s)$ satisfies the Schrödinger differential equation (14) in the adiabatic limit $T\to\infty$; that is the reason why the MS inconsistency appears in their derivation. More importantly, MS inconsistency would lead to vanishing Berry phase [13,19]. However, Berry phase does exist and has been applied in many physical fields, e.g., neutrino oscillations [30], induced gauge field [31], spin hall effect [32] and so on. Because the Schrödinger differential equation can not describe the Berry phase, we conclude that it is not a complete way of describing physical reality [19].

## 6. Conclusion

So far, we have rigorously proved the quantum adiabatic theorem by using the new adiabatic condition, which is given by the theorem 3.2. The traditional adiabatic condition is insufficient and hence can not remove the MS counterexample. The MS inconsistency originates from the use of the Schrödinger differential equation in the derivation of the quantum adiabatic theorem. However, the new adiabatic condition not only removes the MS inconsistency but also rules out the MS counterexample. As a result, we can only reach a complete quantum adiabatic theorem through Schrödinger integral equation rather than Schrödinger differential equation.

# Appendix

## The proof of general Riemann-Lebesgue lemma [29]

Proof. For any $\varepsilon > 0$, we can construct the step function

$$\varphi(x) = \sum_{i=1}^{p} y_i \chi_{[x_{i-1}, x_i)}(x) \text{ with } x \in [a, b], \text{ so that}$$

$$\int_a^b |f(x) - \varphi(x)| dx < \frac{\varepsilon}{2M},$$

where, $\chi_{[x_{i-1}, x_i)}(x) = \begin{cases} 1, x \in [x_{i-1}, x_i) \\ 0, x \notin [x_{i-1}, x_i) \end{cases}$ with $a = x_0 < x_1 < ... < x_p = b$,

and $y_i$ is constant.

If we use the condition (B) and the inequality $\left| \int_a^b \varphi(x) g_n(x) dx \right| \leq \sum_{i=1}^{p} \left| y_i \int_{x_{i-1}}^{x_i} g_n(x) dx \right|$,

then there holds,

$$\left| \int_a^b \varphi(x) g_n(x) dx \right| \leq \frac{\varepsilon}{2} \text{ as } n \geq N,$$

where $N$ is sufficiently large.

Thus, if $n \geq N$ and if we use the condition (A), then there holds

$$\left| \int_a^b f(x) g_n(x) dx \right| \leq \left| \int_a^b [f(x) - \varphi(x)] g_n(x) dx \right| + \left| \int_a^b \varphi(x) g_n(x) dx \right|$$

$$\leq M \int_a^b |f(x) - \varphi(x)| dx + \frac{\varepsilon}{2} < \varepsilon,$$

that is, $\lim_{n \to \infty} \int_a^b g_n(x) f(x) dx = 0$. □

would be, for any function $f(x)$ which is Lebesgue integrable, that $\lim\limits_{n\to\infty}\int_a^b g_n(x)f(x)dx = 0$.

(A) $|g_n(x)| \leq M$ with $x \in [a,b]$, where $M$ is a constant.

(B) For any $c \in [a,b]$, there holds $\lim\limits_{n\to\infty}\int_a^c g_n(x)dx = 0$.

The proof sees Appendix.